\documentclass[%
reprint,
superscriptaddress,
amsmath,
amssymb,aps,
prc
]{revtex4-2}

\usepackage{graphicx}
\usepackage{dcolumn}
\usepackage{bm}
\usepackage{enumitem}
\usepackage{xcolor}
\usepackage{soul}

\begin{document}

\newcommand{\ks}[1]{\textcolor{red}{#1}}
\newcommand{\ig}[1]{\textcolor{blue}{#1}}
\newcommand{\sg}[1]{\textcolor{green}{#1}}

\preprint{APS/123-QED}

\title{Photoneutron reactions on gold in the giant dipole resonance region: \\ reaction cross sections and average kinetic energies of $(\gamma,\,xn)$ photoneutrons}

\author{I.~Gheorghe}\email{ioana.gheorghe@nipne.ro}
\affiliation{National Institute for Physics and Nuclear Engineering,
Horia Hulubei (IFIN-HH), 30 Reactorului, 077125 Bucharest-Magurele, Romania}

\author{T.~Ari-izumi} 
\affiliation{Konan University, Department of Physics, 8-9-1 Okamoto, Higashinada, Kobe 658-8501, Japan}

\author{S.~Goriely} 
\affiliation{Institut d'Astronomie et d'Astrophysique, Universit\'e Libre de Bruxelles, Campus de la Plaine, CP-226, 1050 Brussels, Belgium}

\author{D.~Filipescu} 
\affiliation{National Institute for Physics and Nuclear Engineering,
Horia Hulubei (IFIN-HH), 30 Reactorului, 077125 Bucharest-Magurele, Romania}

\author{S.~Belyshev} 
\affiliation{Lomonosov Moscow State University, Faculty of Physics, 119991 Moscow, Russia}

\author{K.~Stopani} 
\affiliation{Lomonosov Moscow State University, Skobeltsyn Institute of Nuclear Physics, 119991 Moscow, Russia}

\author{H.~Wang} 
\affiliation{Shanghai Advanced Research Institute, Chinese Academy of Sciences, No.99 Haike Road, Zhangjiang Hi-Tech Park, 201210 Pudong Shanghai, China}

\author{G.~Fan} 
\affiliation{Shanghai Advanced Research Institute, Chinese Academy of Sciences, No.99 Haike Road, Zhangjiang Hi-Tech Park, 201210 Pudong Shanghai, China} 

\author{\\H.~Scheit} 
\affiliation{Institut f\"ur  Kernphysik, Technische Universit\"at Darmstadt, Darmstadt, 64289, Germany}

\author{D.~Symochko} 
\affiliation{Institut f\"ur  Kernphysik, Technische Universit\"at Darmstadt, Darmstadt, 64289, Germany}

\author{M. Krzysiek} 
\affiliation{National Institute for Physics and Nuclear Engineering,
Horia Hulubei (IFIN-HH), 30 Reactorului, 077125 Bucharest-Magurele, Romania}
\affiliation{Institute of Nuclear Physics Polish Academy of Sciences, PL-31342 Krakow, Poland}

\author{T.~Renstr\o{}m} 
\affiliation{Department of Physics, University of Oslo, N-0316 Oslo, Norway}
\affiliation{Expert Analytics AS, N-0179 Oslo, Norway}

\author{G.M.~Tveten} 
\affiliation{Department of Physics, University of Oslo, N-0316 Oslo, Norway}
\affiliation{Expert Analytics AS, N-0179 Oslo, Norway}

\author{S.~Miyamoto} 
\affiliation{Laboratory of Advanced Science and Technology for Industry, University of Hyogo, 3-1-2 Kouto, Kamigori, Ako-gun, Hyogo 678-1205, Japan}
\affiliation{Institute of Laser Engineering, Osaka University, 2-6 Yamadaoka, Suita, Osaka 565, Japan}

\author{H. Utsunomiya} 
\affiliation{Konan University, Department of Physics, 8-9-1 Okamoto, Higashinada, Kobe 658-8501, Japan}


\date{\today}

\begin{abstract}
In this work, we present new data on the $^{197}$Au photoneutron reactions in and above the giant dipole resonance region, obtained by using 8 to 39~MeV quasi-monochromatic $\gamma$-ray beams produced at the NewSUBARU facility in Japan and a high-and-flat efficiency neutron detection system. We report absolute cross sections and mean photoneutron energies for the $^{197}$Au$(\gamma,\,inX)$ reactions with $i$~=~1 to 4. The photoabsorption cross section was obtained as the sum of the $(\gamma,\,inX)$ reaction cross sections. The giant dipole resonance parameter values were obtained by fitting the experimental photoabsorption cross sections. The present photoabsorption cross sections are in good agreement with the Saclay results of Veyssiere~\emph{et al.}~\cite{veyssiere_1970}. Thus, our study does not support the recommendation of Berman~\emph{et al.}~\cite{berman_1987} of lowering the Saclay photoabsorption cross sections by 8$\%$. We observed a non-statistical high-energy neutron emission in the $(\gamma,\,n)$ reaction in the low-energy region between $S_n$ and 10~MeV. The present results are compared with data from the literature and statistical model calculations performed with the TALYS and EMPIRE codes. 

\end{abstract}

\maketitle


\section{Introduction} \label{sec_intro}

Photonuclear reactions provide important information on nuclear structure, in particular concerning the giant dipole resonance (GDR), which corresponds to the collective excitation mode dominating the electromagnetic response of atomic nuclei at 10-30 MeV excitation energies. They also represent a powerful tool for characterizing key ingredients in statistical model calculations within the Hauser-Feshbach (HF) formalism, such as the $\gamma$-ray strength function ($\gamma$SF) \cite{goriely_2019} and are of interest to a wide range of applications~\cite{kawano_2020}. 

Many photoneutron reaction cross section studies in the GDR energy range were performed during 1960-1980 at the Saclay and Livermore positron in flight annihilation facilities~\cite{berman_1975}. More recently, in 2015, a campaign of GDR $(\gamma,\,xn)$ experiments was initiated at the NewSUBARU facility using quasi-monochromatic laser Compton scattering (LCS) $\gamma$-ray beams and a dedicated high and flat efficiency neutron detection setup. The first cross section results obtained in the NewSUBARU campaign were included in the Coordinated Research Project (CRP) by IAEA on Photonuclear Data and Photon Strength Functions~\cite{CRP_update_photonucleare}, formed the basis for the PD-2019 evaluations on the investigated nuclei~\cite{kawano_2020} and contributed to solving long standing discrepancies between Saclay and Livermore cross sections.

\begin{figure*}[t]
\centering
\includegraphics[width=0.9\textwidth, angle=0]{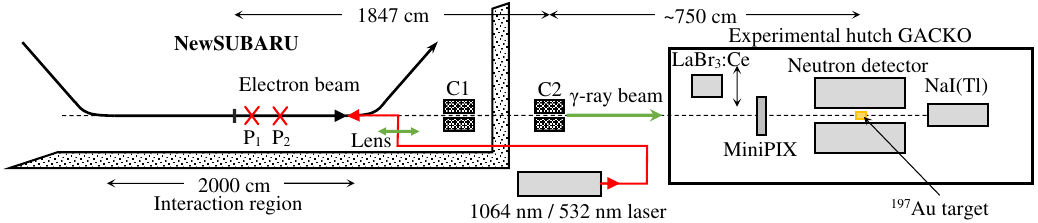}
\caption{Schematic representation (not to scale) of the BL01 LCS $\gamma$-ray beam line and the experimental hutch GACKO at the NewSUBARU facility. P$_1$ and the P$_2$ mark the focus point positions of the 1064~nm and, respectively the 532~nm laser beams at 1.8 and, respectively 3.8~m downstream of the electron beam focus in the middle of the BL01.}\label{fig01_BL01_PRC_Au197_24_crop}     
\end{figure*}

The nucleus $^{197}$Au was part of this set of nuclei, for which the $(\gamma,\,\mathrm{abs})$ and $(\gamma,\,1-4n)$ cross sections measured in the 11 to 39~MeV range were delivered to the IAEA CRP. The photoneutron reaction on $^{197}$Au is of specific interest to nuclear physics applications. As the $^{197}$Au$(\gamma,\,n)$ reaction cross section is suitable for activation measurements, the $(\gamma,\,n)$ reaction on gold is considered a monitor reaction and is thus used to determine the cross sections of most of the reactions of interest for astrophysics, reactor dosimetry and medical applications~\cite{vogt_2002,sonnabend_2004,hasper_2009}. Accurate cross sections for the $^{197}$Au$(\gamma,\,1-3n)$ reactions are also needed to evaluate the competition between the photoneutron reactions on gold and the reactor dosimetry reactions $^{197}$Au$(n,\,2-4n)$ in mixed neutron-$\gamma$ fields from $(\mathrm{n_{th}},\,xn)$ and $(\mathrm{n_{th}},\,x\gamma)$ reactions on uranium nuclei~\cite{CRP_IRDFF_2013,INDC_NDS_0507,capote_2012}.

After the conclusion of the photonuclear IAEA CRP, a more detailed analysis of the NewSUBARU $(\gamma,\,xn)$ data was performed using updated techniques for neutron multiplicity sorting~\cite{gheorghe_2021} and LCS $\gamma$-ray beam spectral distribution~\cite{filipescu_2022,filipescu_2023} and flux~\cite{utsunomiya_2018} characterization. The data set has also been extended with additional low energy $(\gamma,\,n)$ measurements in the vicinity of the neutron emission threshold $S_n$. Results on the $(\gamma,\,\mathrm{abs})$ and $(\gamma,\,1-4n)$ reaction cross sections and on the average kinetic energies of $(\gamma,\,1-4n)$ photoneutrons in the extended 8 to 39~MeV energy range are reported in this paper.

In Sect.~\ref{sec_exp_method}, we present the experimental technique and methodology, with focus on diagnostics of the incident LCS $\gamma$-ray beams and neutron detection. The data analysis methods are discussed in Sect.~\ref{sec_data_analysis}. Results are discussed and compared with preceding data in Sect.~\ref{sec_results} and with theoretical calculations in Sect.~\ref{sec_STAT_calc}. A summary and conclusions are given in Sect.~\ref{sec_summary}.

\section{Experimental method} \label{sec_exp_method}

Quasi-monoenergetic photon beams were produced at the laser Compton scattering $\gamma$-ray source of the NewSUBARU facility~\cite{amano_2009,horikawa_2010} at 54 energies in the 8.59 to 39.32~MeV energy range. 
Laser photons were backscattered of relativistic electrons circulating in the NewSUBARU 0.5 to 1.5~GeV electron storage and accelerator ring with currents between 60 and 300~mA. After passing through a set of two 10-cm-long lead collimators with apertures of 3 and 2~mm, the 4~mm diameter pencil-like photon beam~\cite{Ariizumi_2023} entered the GACKO experimental hutch (Gamma collaboration hutch of Konan University) and irradiated the gold target placed in the center of the neutron detection array. Figure~\ref{fig01_BL01_PRC_Au197_24_crop} shows a diagram of the LCS $\gamma$-ray beam line and of the experimental hutch. 

The energy resolution and flux of the $\gamma$-ray beam were monitored with a lanthanum bromide (LaBr$_3$:Ce) and a NaI:Tl detector, respectively. The reaction neutrons were moderated and recorded by $^3$He counters placed in concentric rings around the target. The data were written in a triggerless list mode, using an eight-parameter 25 MHz digital data acquisition system which recorded the time and energy of the signals provided by the LaBr$_3$:Ce and NaI detectors, the neutron detection time in each ring of $^3$He counters and the external laser triggering signal. 

We measured the $\sigma_{inX}$ sum cross sections of all photoneutron reactions with $i$ neutrons in the final state, accompanied or not by charged particle emission:
\begin{align}
\sigma_{inX}          & \equiv \sigma_{\gamma,\,inX} \nonumber \\
                      & \equiv \sigma(\gamma,\,inX) \nonumber \\
                      & = \sigma(\gamma,\,in) + \sigma(\gamma,\,inp) + \sigma(\gamma,\,in\alpha) + \dots       
\end{align}
However, the large Coulomb barrier in $^{197}$Au~($\sim$25~MeV for $\alpha$ particles) hinders the emission of charged particles. Thus, for incident photon energies covering the GDR range up to $\sim$20~MeV, both the cross sections and the average photoneutron energies for the $(\gamma,\,inX)$ reactions generally coincide with or are a good approximation to the $(\gamma,\,in)$ neutron emission only cross sections~\cite{kawano_2020}. 

\subsection{LCS $\gamma$-ray beams}

\subsubsection{LCS $\gamma$-ray beam energy}

The maximum energy of the LCS $\gamma$-ray beam is directly determined by the laser wavelength and by the electron beam energy, which has been calibrated with the accuracy on the order of 10$^{-5}$ \cite{utsunomiya_2014}. Solid state lasers with 1064 and 532~nm wavelength were used in the present experiment. The LCS $\gamma$-ray beam energy was finely tuned by setting the electron beam energy between approximately 570 and 1070~MeV. Table~\ref{table_beams} gives the precise electron beam energy ranges used in connection with each of the two lasers and the resulting LCS $\gamma$-ray beam energies. 

\begin{table*}
\caption{\label{table_beams} Parameters for the laser, electron and LCS $\gamma$-ray beams, $^{197}$Au samples and type of neutron detector used. $E_{e^-}$ is the electron beam energy and $E_{m}$ the maximum energy of the $\gamma$-ray beam~\cite{utsunomiya_2014}. $t$ is the target thickness and $\mathcal{T}$ the photon transmission through the target.}
\begin{ruledtabular}
\begin{tabular}{c|ccccccccccc}
type of         & neutron  & $\lambda$ & $f_\mathrm{laser}$ & $t$  & $E_{e^-}$    & $E_\mathrm{m}$ & $\mathcal{T}$ & Mean nb. of    & Incident flux &  \\ 
measurement     & detector & (nm)      & (kHz)              & (mm) & (MeV)        & (MeV)          & ($\%$)        & $\gamma$/pulse & ($\times 10^4$ $\gamma/$s) &  \\ \hline              
neutron         & HED      & 1064      & 20                 & 4    & 697.5--897.5 & 8.59--14.17    & 67--70        & 3--7           & 4.8--11.2 &  \\      
counting        & FED      & 532       & 20                 & 2    & 570.7--641.2 & 11.42--14.37   & 81--82        & 2--3           & 3.2--4.8  &  \\  \hline     
multiplicity sorting   & FED      & 532       & 1                  & 4    & 652.3--1068.7& 14.87--39.32   & 57--65        & 7--25          & 0.5--2.0  &  \\   
\end{tabular}
\end{ruledtabular}
\end{table*}

\subsubsection{Time structure and incident photon flux}

The pulsed time structure of the LCS $\gamma$-ray beam followed the slow time structure of the laser beam (1~--~20~kHz, 20~--~40~ns pulse width) and the fast structure of the electron beam (500~MHz, 60~ps pulse width). In photoneutron experiments with slow response neutron detectors, by $\gamma$ pulse we refer to the 1-20 kHz frequency pulse determined by the temporal structure of the laser. For measurements below the two neutrons separation energy (S$_{2n}$), the laser was operated at a frequency of 20~kHz, corresponding to an interval of 50~ms between two consecutive LCS $\gamma$-ray beam pulses. For neutron multiplicity sorting measurements above S$_{2n}$, we used a reduced frequency of 1 kHz, with an interval of 1~ms between two consecutive gamma pulses, which is of the order of the neutron die away time in the moderated detection system. A 100~ms macro-time structure of alternating 80~ms beam-on followed by 20~ms beam-off intervals was used for background monitoring and rejection. 

At the NewSUBARU LCS $\gamma$-ray source, the number of photons in a $\gamma$-ray pulse follows the Poisson distribution~\cite{kondo_2011}. The incident photon flux was obtained by applying the pile-up or Poisson fitting method~\cite{utsunomiya_2018} on the response functions of the flux monitor NaI(Tl) detector (8" diameter $\times$ 12" length). The mean numbers of $\gamma$ rays per pulse and the incident photon flux values are given in Table~\ref{table_beams}.

\subsubsection{Spectral distribution}

\begin{figure}[t]
\centering
\includegraphics[width=0.98\columnwidth, angle=0]{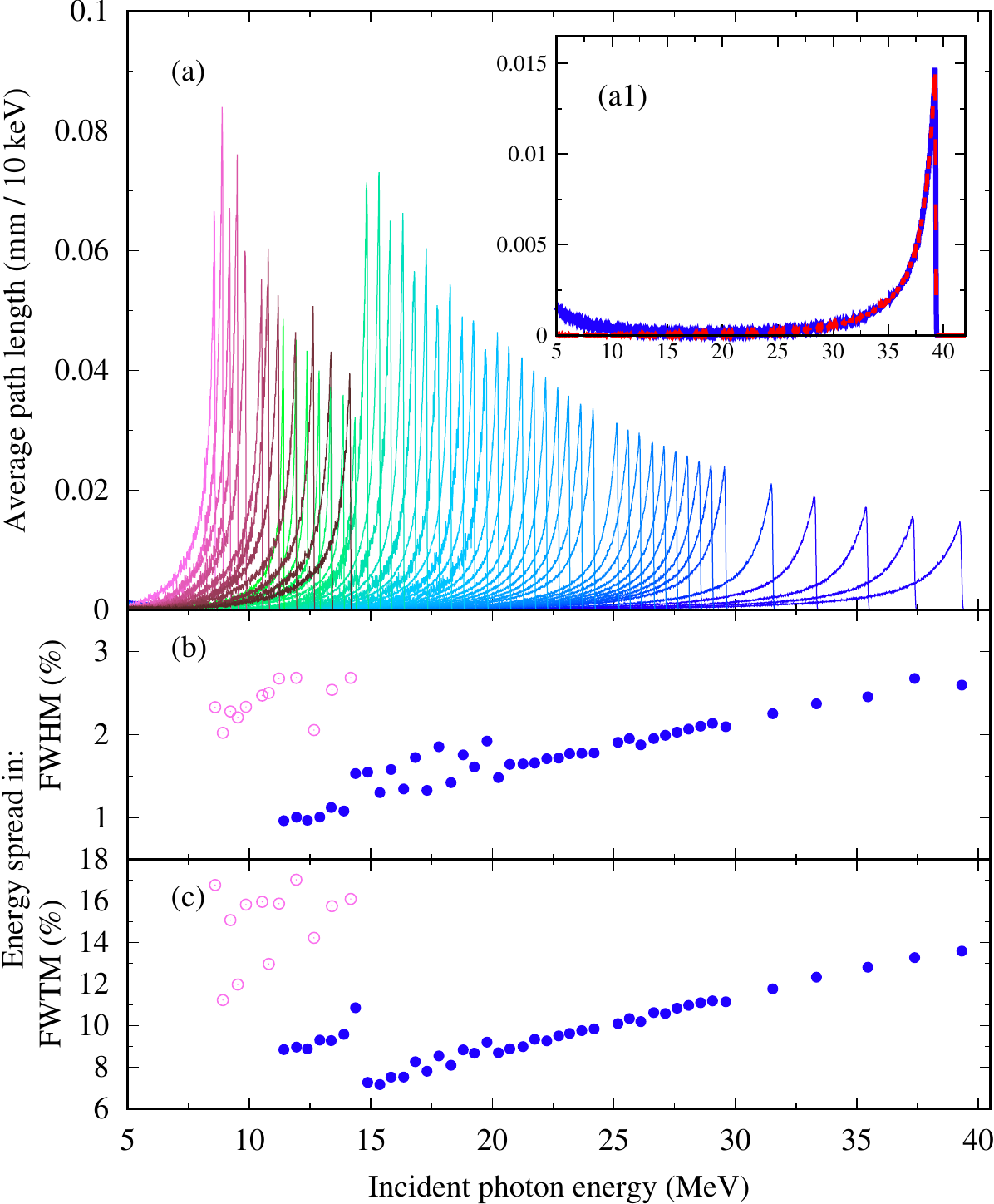} 
\caption{(a) The simulated energy profiles of the 54 LCS $\gamma$-ray beams used in the present experiment represented by the $L(E_\gamma,E_\mathrm{m})$ distributions which account for the electromagnetic interaction of the incident photon beam with the target material. The $\gamma$ beams obtained with the 1064~nm wavelength laser are shown in magenta and brown, the ones obtained with the 532~nm laser in green and blue. The (a1) inset shows the incident spectrum (dashed red) and the $L(E_\gamma,E_\mathrm{m})$ (blue) for the maximum energy 39.32~MeV beam. The energy spread of the LCS $\gamma$-ray beams is shown in FWHM in (b) and in FWTM in (c): magenta for the 1064~nm laser, blue for the 532~nm laser.}\label{fig02_inc_spectra_lin}
\end{figure}

The energy spectra of the 54~photon beams used in the present experiment are shown in Fig.~\ref{fig02_inc_spectra_lin}. They are obtained from the comparison of Monte Carlo simulations of the 3.5" diameter $\times$ 4.0" length LaBr$_3$:Ce monitor detector pulse-height spectra to the measured pulse-height spectra. The simulations were performed with the realistic LCS $\gamma$-source simulation code $\textsc{eliLaBr}$~\cite{eliLaBr_github,filipescu_2023,filipescu_2022}, which incorporates the $\gamma$-ray beam production in the 20~m long interaction region, its transport through the double collimation system and its interaction with the LaBr$_3$:Ce monitor detector.

In the present study, LCS $\gamma$-ray beams with sufficiently high energies were used so that the secondary particles generated by the electromagnetic interaction of the incident beam with the target material had energies higher than the neutron emission threshold. Therefore, the contribution of these secondary photons to the measured reaction yields had to be considered.

Using the $\textsc{Geant4}$ package, we simulated the electromagnetic interaction of the incident LCS $\gamma$-ray beams obtained with the $\textsc{eliLaBr}$ code with the target material. The resulting spectra were represented using the $L(E_\gamma,E_\mathrm{m})$ distributions defined as the average path length per unit energy traveled through the target by an $E_\gamma$ photon in an LCS $\gamma$-ray beam of $E_\mathrm{m}$ maximum energy~\cite{filipescu_2023}. The $L(E_\gamma,E_\mathrm{m})$ distributions account for the $\gamma$-beam self-attenuation in the irradiated sample material and for the secondary radiation generated by electromagnetic interaction of the $\gamma$-beam with the target.

The spectral distributions of the LCS $\gamma$-ray beams used in the present experiment are shown by $L(E_\gamma,E_\mathrm{m})$ histograms in Fig.~\ref{fig02_inc_spectra_lin}(a). The low energy magenta and brown spectra shown were obtained using the 1064~nm laser and the green and blue ones with the 532~nm laser. A comparison between the energy spectrum of the incident LCS $\gamma$-ray beam (red) and the $L(E_\gamma,E_\mathrm{m})$ spectral distribution generated by $\gamma$-beam electromagnetic interaction with the sample material (blue) is shown in the (a1) inset of Fig.~\ref{fig02_inc_spectra_lin} for the highest energy setting of 39.32~MeV. Both distributions show the characteristic low energy tail which decreases starting from the maximum energy edge. The contribution of secondary photons in the $L(E_\gamma,E_\mathrm{m})$ distribution is shown by the rise in the low energy tail starting at $\sim$17~MeV. We can notice that, for $\sim$30--40~MeV photon beams, there is a non-negligible share of secondary photons with sufficiently high energies to induce nuclear reactions in the target.

Figures~\ref{fig02_inc_spectra_lin}(b) and (c) show the energy spread of the incident LCS $\gamma$-ray beams expressed by the full width at half maximum (FWHM) and by the full width at one tenth maximum (FWTM), respectively. The FWHM varied between 2 and 3~$\%$ for the low energy beams generated using the 1064~nm laser and between 1 and 2.5~$\%$ for the higher energy ones generated with the 532~nm laser. Due to the low energy tail of the distributions, the FWTM is significantly larger, with values between 7 and 17\%.

\subsection{Targets}

Two disks of $^{197}$Au (99.95~$\%$ chemical purity) were used in the present experiment. The same samples were used in the previous measurements of Itoh \emph{et al.}~\cite{itoh_2011}. Each disk had a 2~mm thickness with 10~$\mu$m estimated uncertainty and 20~mm diameter. Measurements were performed with the stacked two disks or with one disk, as given in Table~\ref{table_beams} for each incident energy range. The photon transmission through the gold samples is also listed in Table~\ref{table_beams}.

\subsection{Neutron detection}

\begin{figure}[t]
\centering
\includegraphics[width=0.98\columnwidth, angle=0]{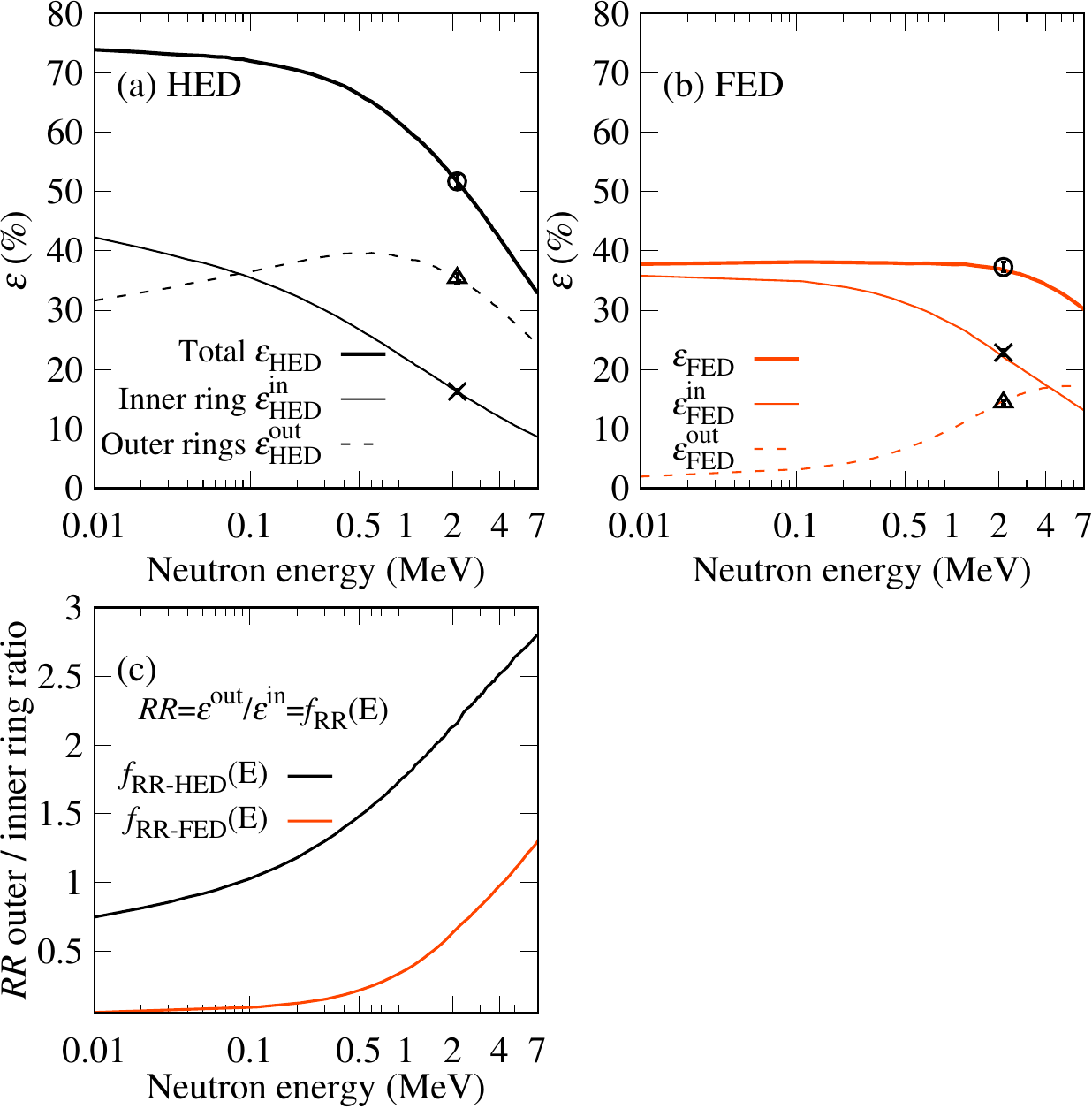} 
\caption{Calibration of neutron detectors. Experimental $^{252}$Cf measurements for the total (circle), inner ring (cross) and summed outer rings (triangle) detection efficiencies reproduced by \textsc{mcnp}~\cite{mcnp} simulations for (a) the HED and (b) the FED detectors by considering e\-vap\-o\-ra\-tion neutron spectra. (c) The $\varepsilon^\mathrm{in}/\varepsilon^\mathrm{out}$ ratio between the efficiencies of the inner and summed outer rings of counters, here referred to as ring ratio functions, and computed for the HED (black) and FED (orange) detectors, as function of the average neutron energy.}\label{fig_fed_eff_new}     
\end{figure}

Neutrons emitted from the irradiated gold targets were detected with $^3$He detectors placed in concentric rings in moderator material. We used two different three-ring detection geometries:
\begin{itemize}
\item the High Efficiency Detector (HED) developed for low cross-section $(\gamma,\,n)$ measurements in the vicinity of $S_n$ and described in Ref.~\cite{utsunomiya_2015}, for which the total detection efficiency represented in Fig.~\ref{fig_fed_eff_new}(a) varies between 80 and 30 $\%$ over a range of neutron energies between 10 keV and 5 MeV. 
\item the Flat Efficiency Detector (FED) developed for $(\gamma,\,xn)$ neutron multiplicity sorting measurements and described in Refs.~\cite{utsunomiya_2017,gheorghe_2021}, for which the total detection efficiency represented in Fig.~\ref{fig_fed_eff_new}(b) varies between 38 and 33 $\%$ over a range of neutron energies between 10 keV and 5 MeV. 
\end{itemize}
Based on the reasonable assumption that the dominant decay mode of the GDR in the mildly deformed heavy nucleus $^{197}$Au is statistical with a direct neutron emission contribution estimated to be less than 2$\%$~\cite{orlin_2023}, neutron evaporation spectra were considered for both calibrations represented in Fig.~\ref{fig_fed_eff_new}. The efficiency value is represented at the average energy of the neutron spectra.

Table~\ref{table_beams} lists all combinations between detector types, lasers and target thicknesses used in the experiment. The measurements carried out below $S_{2n}$ by operating the laser at the frequency of 20~kHz were dedicated to the direct counting of the neutrons emitted in the $^{197}$Au$(\gamma,\,n)$ reaction. The experimental method is described in Refs.~\cite{filipescu_2014,utsunomiya_2015}.

The measurements carried out above $S_{2n}$ by operating the laser at the frequency of 1~kHz were dedicated to neutron multiplicity sorting. We separately picked events in which 1, 2, 3,~$\dots$ neutrons were recorded in the entire detection system between two consecutive $\gamma$-beam pulses, and which we refer to as $i$-fold neutron coincidence events. The subtraction of the neutron background is based on the analysis of the time moderation spectra of the neutrons in the FED and has been discussed in detail in Refs.~\cite{utsunomiya_2017,gheorghe_2021,gheorghe_2017,filipescu_2024,gheorghe_2024}. 

\section{Data analysis} \label{sec_data_analysis}

\subsection{Number of neutron coincidence events and $i$-fold cross sections} \label{sec_Ni_def} 

The primary experimental information in neutron multiplicity sorting experiments are the numbers of $i$-fold coincidence events $n_i(E_\mathrm{m})$ measured for a given incident LCS $\gamma$-ray beam of maximum energy $E_\mathrm{m}$. To have a convenient graphical representation of $n_i$, we further use the $i$-fold neutron cross sections $N_i(E_\mathrm{m})$ defined as
\begin{equation}
N_i (E_\mathrm m) = \cfrac{n_i(E_\mathrm m) }{N_\gamma (E_\mathrm m) n_\mathrm T \xi(E_\mathrm m) } 
\end{equation} 
where $n_\mathrm T$ is the concentration of target nuclei and $N_\gamma$ is the incident photon number. $\xi=[1-\mathrm{exp}(-\mu L)]/\mu$ is a thick target correction factor given by the target thickness $L$ and attenuation coefficient $\mu$.

\begin{figure*}[t]
\centering
\includegraphics[width=0.98\textwidth, angle=0]{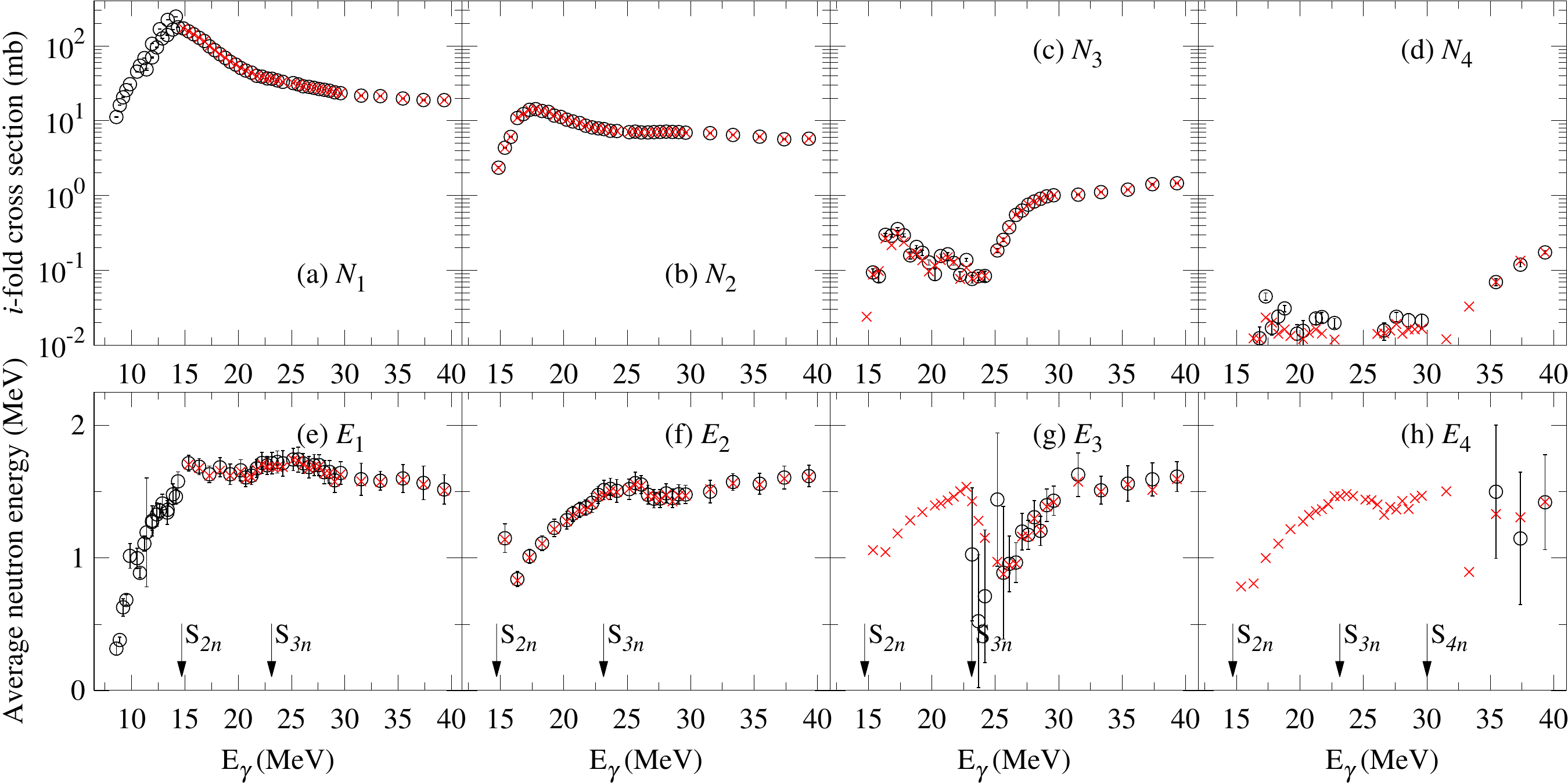} 
\caption{Direct observables in the $^{197}$Au study: (a--d) $i$-fold cross sections and (e--h) average neutron energies of neutrons recorded in $i$-fold coincidence events. Experimental values are shown by open black circles and best fit ones obtained through the multiple firing neutron multiplicity sorting method discussed in Sec.~\ref{sec_NMS} by red crosses. The error bars for the average energies of $i$-fold coincidence events represent the statistical component and a 3$\%$ systematic component accounting for the uncertainty in the neutron detection efficiency calibration while the ones for $i$-fold cross sections $N_i$ are statistical only.}\label{fig_nev_rate}     
\end{figure*}

Figures~\ref{fig_nev_rate}(a--d) show the experimental $i$-fold cross sections (open black circles) for the $^{197}$Au$(\gamma,\,inX)$ reactions with $i$ ranging from (a) 1 to (d) 4. We notice that $N_2$, $N_3$ and $N_4$ $i$-fold cross sections rise at the opening of the 
$(\gamma,\,2n)$, $(\gamma,\,3n)$ and $(\gamma,\,4n)$ channels, respectively, 
with small non-zero contributions below the corresponding thresholds. 
This shows that the experiment was conducted in close to single-firing conditions, 
with only small probability for multiple reactions being induced in the target by the same $\gamma$-ray pulse. 
The treatment of the fraction of multiple-firing neutron events will be discussed in Sect.~\ref{sec_NMS}.   

\subsection{Average energies of neutrons recorded in $i$-fold coincidence events} \label{sec_Ei}

Information on the average energy of the neutron emission spectra was extracted by applying the ring-ratio method described in detail in Refs.~\cite{filipescu_2024,gheorghe_2024}. The method is based on the calculation of the theoretical ratios between the detection efficiencies of the outer and inner neutron rings $f_\mathrm{RR-HED}$ and $f_\mathrm{RR-FED}$ for the HED and FED detectors, respectively, which are shown in Fig.~\ref{fig_fed_eff_new}(c). The average energy of neutrons recorded in $i$-fold coincidence events $E_i$ was determined by evaluating on the corresponding HED or FED theoretical ratio curve the experimental ratios between the numbers of neutrons recorded in $i$-fold events in the outer and inner rings $n^\mathrm{out}_i(E_\mathrm m)$ and $n^\mathrm{in}_i(E_\mathrm m)$, respectively
\begin{equation}
E_i(E_\mathrm m)= f_\mathrm{RR-FED/HED}^{-1} \left( \cfrac{n^\mathrm{out}_i(E_\mathrm m)}{n^\mathrm{in}_i(E_\mathrm m)} \right).
\end{equation}
Figures~\ref{fig_nev_rate}(e--h) show the experimental ring-ratio extracted average energies $E_i$ (open black circles) of neutrons recorded in $i$-fold events for i~=~1~(e) to i~=~4~(h).  

\subsection{Neutron multiplicity sorting} \label{sec_NMS}

Let us consider a photon beam of energy $E_\gamma$ incident on the $^{197}$Au target. For $S_{\texttt{x}n}<E_\gamma<S_{(\texttt{x}+1)n}$, the $(\gamma,\,inX)$ reactions with $i$ taking values from 1 to $\texttt{x}$ can be induced in the target, each characterized by the $\sigma_{inX}$ cross section and $E_{inX}$ average energy of the photoneutron emission spectrum. The $S_{in}$ neutron separation energies with $i$ from 1 to 5 are given for $^{197}$Au in Table~\ref{table_Sxn}.

\begin{table}[b]
\caption{\label{table_Sxn} Threshold energies $S_{in}$ for the emission of $i$ neutrons from the $^{197}$Au nucleus. All values are given in MeV.}
\begin{ruledtabular}
\begin{tabular}{lccccc}
   \textrm{S$_n$} & \textrm{S$_{2n}$}& \textrm{S$_{3n}$} & \textrm{S$_{4n}$} & \textrm{S$_{5n}$} \\ \colrule
           8.073  &  14.716          & 23.143            & 30.023            & 38.728 \\
\end{tabular}
\end{ruledtabular}
\end{table}

\begin{figure*}[t]
\centering
\includegraphics[width=0.98\textwidth, angle=0]{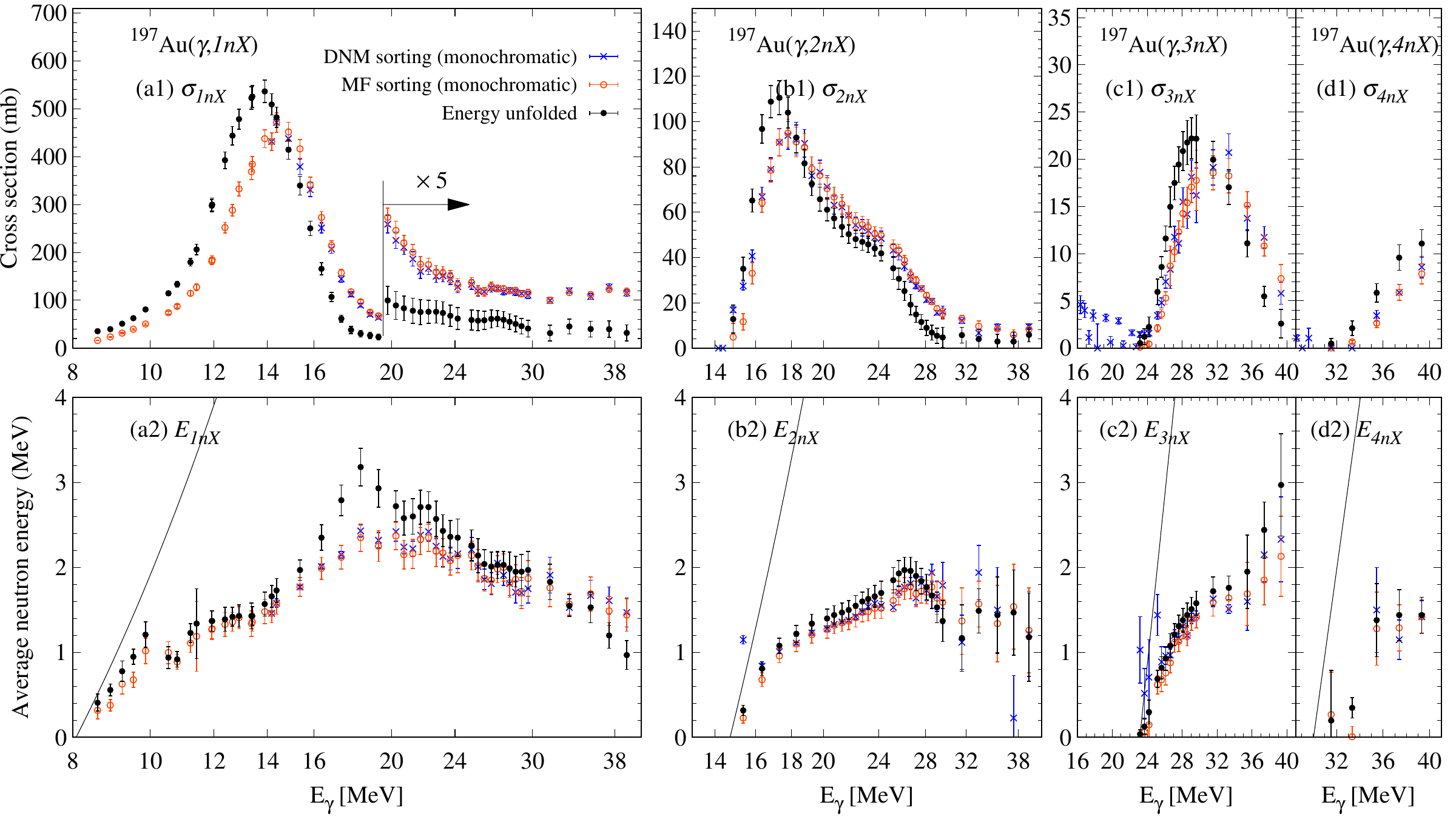} 
\caption{$^{197}$Au$(\gamma,\,inX)$ results: [upper figures (a1)--(d1)] reaction cross sections and [lower figures (a2)--(d2)] average energies of photoneutrons. The results of the DNM sorting method are shown by empty red dots and the multiple firing corrected ones by blue crosses. The final results, obtained after correcting for the spectral distribution of the LCS $\gamma$-ray beams are shown by full black dots. The error bars account for the total uncertainty. The solid lines correspond to the maximum neutron energies given by kinematics $196/197\cdot(E_\gamma-S_{in})$. The (a) and (b) horizontal axes are in log scale.}\label{fig_mono_vs_unfolded}   
\end{figure*}

For (i) a nonunity efficiency neutron detection system and (ii) single firing conditions, in which no more than one reaction is induced in the target by each LCS $\gamma$-ray beam bunch, it follows that the direct observables $N_i$ and $E_i$ are given by contributions from all reaction channels $(\gamma,\,knX)$ with $k\ge i$. Considering also the energy independent detection efficiency of the FED, one can apply the direct neutron multiplicity (DNM) sorting method to discriminate the contributions of the competing $(\gamma,\,inX)$ reactions.  

However, given that the photon multiplicity in the 1~kHz LCS $\gamma$-ray pulses was Poisson distributed with mean values between 7 and 25, as given in Table~\ref{table_beams}, there was a small probability for more than one reaction to be induced in the target by the same photon bunch. The small presence of these multiple-firing events in the $^{197}$Au data set can be noticed by the nonzero $N_3$ and $N_4$ values below $S_{3n}$ and $S_{4n}$, respectively. 

Thus, we applied the multiple firing neutron multiplicity sorting method described in Refs.~\cite{gheorghe_2021,filipescu_2024,gheorghe_2024}, which is a statistical treatment that models the firing of all combinations of energetically allowed $(\gamma,\,inX)$ reactions for each incident energy value and the subsequent coincident detection of the emitted neutrons. Using a minimization procedure, we iteratively adjusted the free parameters -- the $(\gamma,\,inX)$ cross sections and average neutron energies, until the calculated $i$-fold cross sections $N_i^\mathrm{MF}$ and average energies $E_i^\mathrm{MF}$ reproduced the experimental observables. Figure~\ref{fig_nev_rate} shows that the experimental $N_i$ and $E_i$ (open black circles) are well reproduced by the $N_i^\mathrm{MF}$ and $E_i^\mathrm{MF}$ results of the $multiple$-firing procedure (red crosses) on the entire excitation energy region and for all observed neutron multiplicities.

Figure~\ref{fig_mono_vs_unfolded} shows in (a1)--(d1) the $^{197}$Au$(\gamma,\,inX)$ cross sections and in (a2)--(d2) the average energies of the photoneutron spectra for which the experimental observables $N_i$ and $E_i$ were best reproduced within the multiple firing neutron multiplicity sorting procedure (empty red dots). For comparison, we also show the results of the DNM sorting method (blue crosses). Below $S_{2n}$, only the $(\gamma,\,n)$ channel contributes to the recorded events and no multiplicity sorting is needed, thus only the DNM results are shown in Fig.~\ref{fig_mono_vs_unfolded}. Above $S_{2n}$, we notice that the multiple firing sorting procedure brings small corrections to the DNM results both for the cross sections and average neutron energies. As observed also in the case of $^{159}$Tb~\cite{gheorghe_2017}, $^{238}$U, $^{232}$Th~\cite{filipescu_2024} and $^{208}$Pb~\cite{gheorghe_2024}, multiple firings affect most significantly the results for a weak reaction channel in competition with a much stronger one, such as the $^{197}$Au$(\gamma,\,2nX)$ ones in competition with $^{197}$Au$(\gamma,\,1nX)$ just above $S_{2n}$.

\subsection{Correction for the spectral distribution of the incident LCS $\gamma$-ray beams} \label{sec_energy_unfolding}

The results of the neutron multiplicity sorting procedure discussed above must be further corrected by considering the spectral distribution of the incident LCS $\gamma$-ray photon beams. 

First, we related the uncorrected $^{197}$Au$(\gamma,\,inX)$ cross sections $\sigma_{inX}^\mathrm{MF}(E_\mathrm m)$ and average neutron energies $E_{inX}^\mathrm{MF}(E_\mathrm m)$ to the folding between the beam spectral distribution $L(E_\gamma,E_\mathrm m)$ and the true, $\gamma$-ray energy dependent quantities $\sigma_{inX}(E_\gamma)$ and $E_{inX}(E_\gamma)$, respectively:
\begin{align} \label{eq_folded_cs}
\sigma_{inX}^\mathrm{MF}(E_\mathrm m) & = \cfrac{1}{\xi(E_\mathrm m)} \int_0^{E_\mathrm m} L(E_\gamma,E_\mathrm m) \sigma_{inX}(E_\gamma)\,dE_\gamma \\
E_{inX}^\mathrm{MF}(E_\mathrm m) & = \cfrac{\int_0^{E_\mathrm m} \!\! E_{inX}(E_\gamma) L(E_\gamma,\!E_\mathrm m)\sigma_{inX}(E_\gamma)dE_\gamma}{\sigma_{inX}^\mathrm{MF}(E_\mathrm m)\xi(E_\mathrm m)}. \label{eq_folded_en_gxn}
\end{align}

Second, we applied an iterative unfolding method described in Refs.~\cite{Renstrom18,LarsenTveten23}, in which a trial function is iteratively adjusted bin by bin based on the difference between its folding with the spectral distribution and the experimental uncorrected quantity. The method makes no a priori assumptions on the shape of the corrected, energy unfolded functions. We applied it first on the $^{197}$Au$(\gamma,\,inX)$ cross sections and then on the average neutron energies, independently for each $i$~=~1 to 4 multiplicity. 

Figure~\ref{fig_mono_vs_unfolded} shows the cross sections and the average neutron energies after applying the correction for the spectral distribution of the incident LCS $\gamma$-ray beam (filled black circles) in comparison with the uncorrected results. 
The statistical uncertainty in the number of detected neutron events of 1 to 4 multiplicities along with uncertainties of 3$\%$ in the neutron detection efficiency~\cite{utsunomiya_2017,gheorghe_2017}, 3$\%$ for the photon flux determination and 0.5$\%$ for the target thickness are included in the error bars. An estimate for the level of uncertainty in the unfolding procedure is obtained by defining upper and lower limits for the monochromatic cross sections by adding and subtracting, respectively, the contribution given by the statistical uncertainty in the number of neutron events. The upper and lower limits are unfolded separately~\cite{Renstrom18,LarsenTveten23}. As discussed in Ref.~\cite{gheorghe_2021}, depending on the incident energy range and neutron emission multiplicity of the reaction, the dominant contribution to the total uncertainty varies between the statistical component given by the recorded neutron event numbers and the uncertainty in the calibration of the neutron detector efficiency.

\section{Experimental results}\label{sec_results}

\subsection{Photoabsorption cross sections}

\begin{figure}[t]
\centering
\includegraphics[width=0.98\columnwidth, angle=0]{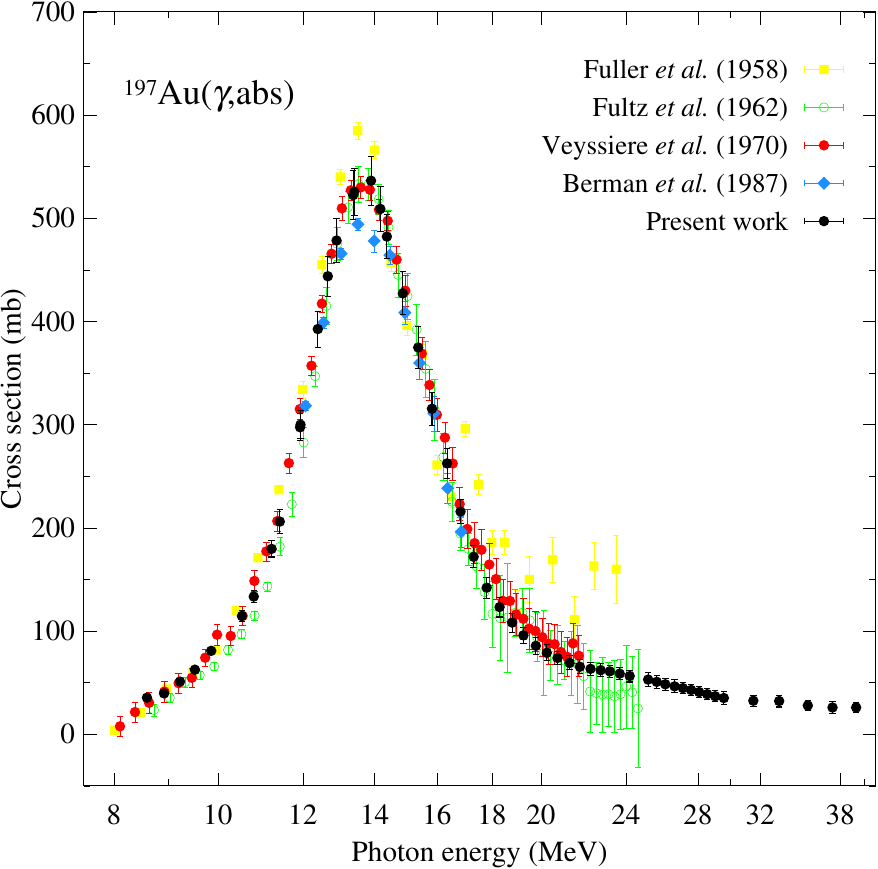} 
\caption{Present photoabsorption cross sections for $^{197}$Au compared with positron annihilation data measured at Saclay by Veyssiere~\emph{et al.}~\cite{veyssiere_1970} and at Livermore by Fultz \emph{et al.}~\cite{fultz_1962} and by Berman \emph{et al.}~\cite{berman_1987}, and with the older bremsstrahlung results of Fuller~\emph{et al.}~\cite{fuller_1958}.}\label{fig06_cs_abs}     
\end{figure}

Figure~\ref{fig06_cs_abs} shows the present $^{197}$Au photoabsorption cross sections compared with existing positron annihilation in flight results obtained at Saclay~\cite{veyssiere_1970} and Livermore~\cite{fuller_1958,fultz_1962,berman_1987}. Both the present and the existing results were obtained in direct neutron detection experiments by summing the partial $(\gamma,\,inX)$ photoneutron cross sections:
\begin{equation}
\sigma(\gamma,\,\mathrm{abs}) \equiv \sum_i \sigma(\gamma,\,inX).
\end{equation}

There is good agreement between the present photoabsorption cross sections and the Saclay results of Veyssiere~\emph{et al.}~\cite{veyssiere_1970}. Thus, the present results do not support the recommendation of Berman~\emph{et al.}~\cite{berman_1987} of lowering the Saclay photoabsorption cross sections by 8$\%$. The present results are also in good agreement with the Livermore data of Fultz~\emph{et al.}~\cite{fultz_1962}, with the notable exception of the increasing slope of the GDR peak, where the Fultz data are systematically lower than all the other experimental data sets. Also, on the high-energy tail of the GDR peak, at excitation energies of about 18~MeV, the present photoabsorption cross sections show a faster descent than both the Saclay data of Veyssiere~\emph{et al.}\cite{veyssiere_1970} and the Livermore ones of Fultz~\emph{et al.}~\cite{fultz_1962}. The present experiment extends the investigated energy range above the current limit of 24~MeV excitation energy. We notice a broad and low shoulder centered at 25~MeV, above the GDR peak.

\subsection{Photoneutron cross sections}

\begin{figure}[t]
\centering
\includegraphics[width=0.98\columnwidth, angle=0]{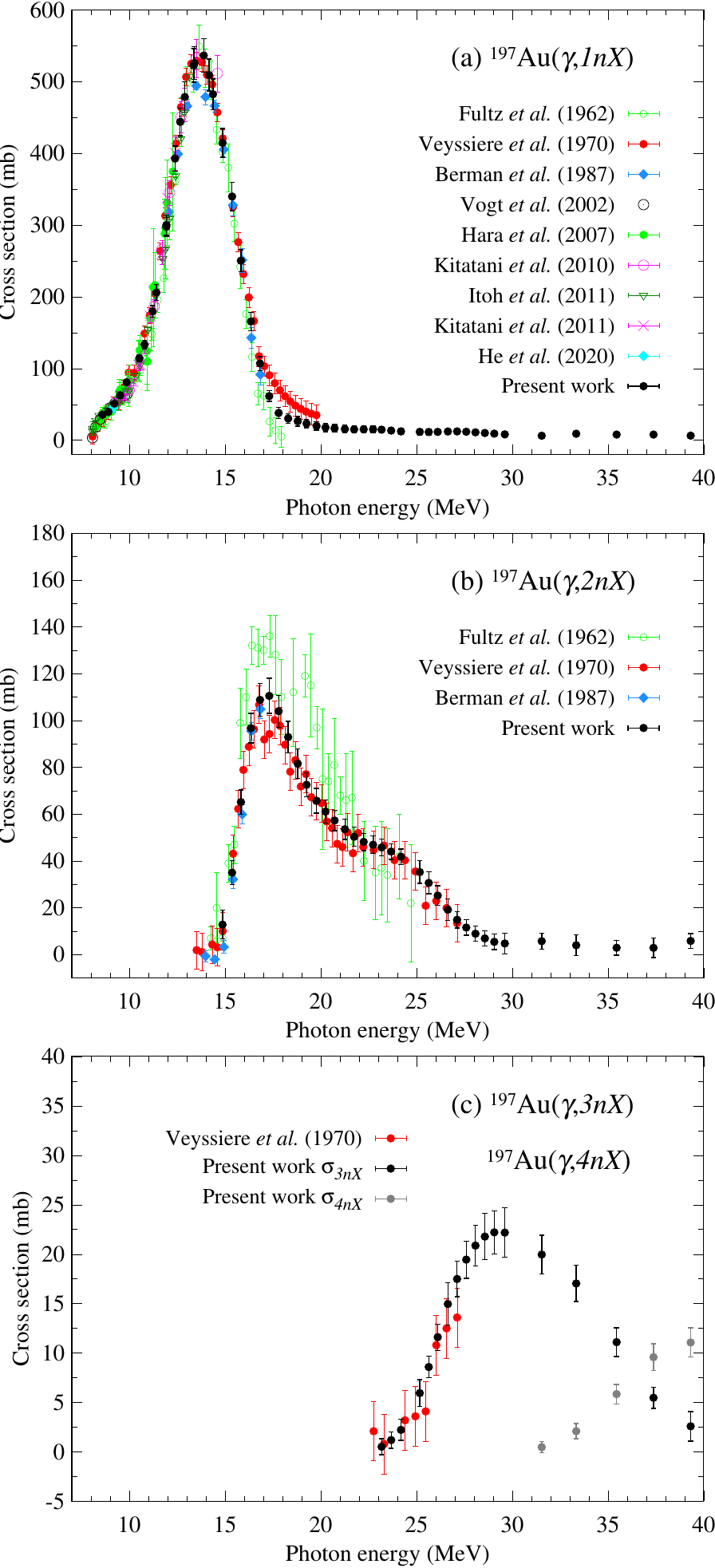} 
\caption{Present photoneutron cross sections for the (a)~$(\gamma,\,1nX)$, (b)~$(\gamma,\,2nX)$, (c)~$(\gamma,\,3nX)$ and $(\gamma,\,4nX)$ reactions in $^{197}$Au compared with existing data.
}\label{fig07_cs_gxn}     
\end{figure}

Figure~\ref{fig07_cs_gxn}(a) shows the present cross sections for the $^{197}$Au$(\gamma,\,1nX)$ reaction in comparison with the existing data. The positron in flight annihilation data of Veyssiere~\emph{et al.}~\cite{veyssiere_1970}, Fultz \emph{et al.}~\cite{fultz_1962} and Berman~\emph{et al.}~\cite{berman_1987} as well as the more recent LCS $\gamma$-ray beam data of Hara~\emph{et al.}~\cite{hara_2007}, Kitatani~\emph{et al.}~\cite{kitatani_2010,kitatani_2011} and Itoh~\emph{et al.}~\cite{itoh_2011} were obtained in direct neutron detection experiments, and thus represent $\sigma_{1nX}$ cross sections similar to the present results. The bremmstrahlung data of Vogt~\emph{et al.}~\cite{vogt_2002} and the 9.17~MeV measurement performed by He~\emph{et al.}~\cite{chuangyeHe_2020} using the $^{13}$C$(p,\,\gamma)$ reaction were obtained by measuring the activation on the ground state of the residual $^{196}$Au. 

On the low energy increasing slope of the $(\gamma,\,n)$ excitation function there is overall good agreement between the present results and the existing ones. As for the photoabsorption case, in the peak region and on the decreasing slope spanning between 13 and 16~MeV excitation energy, the present $\sigma_{1nX}$ cross sections are in good agreement with the Saclay results~\cite{veyssiere_1970} and with the Livermore ones of Fultz~\emph{et al.}~\cite{fultz_1962}. However, on the high energy tail of the $(\gamma,\,n)$ excitation function, the present cross sections are systematically below the Saclay ones and above the Fultz ones. For excitation energies above 20~MeV, the $\sigma_{1nX}$ cross sections show a continuous and slow decrease from 16 to 7~mb. 
 
Figure~\ref{fig07_cs_gxn}(b) shows the present cross sections for the $^{197}$Au$(\gamma,\,2nX)$ reaction in comparison with the positron in flight annihilation data of Veyssiere~\emph{et al.}~\cite{veyssiere_1970}, Fultz \emph{et al.}~\cite{fultz_1962} and Berman~\emph{et al.}~\cite{berman_1987}. Here, the present cross sections are slightly above the ones of Veyssiere on the peak cross sections and systematically below the Fultz ones. Considering that for the $(\gamma,\,1nX)$ reaction is the other way round, the differences may originate from the neutron multiplicity sorting procedures used in the data analysis. 

Figure~\ref{fig07_cs_gxn}(c) shows the present cross sections for the $(\gamma,\,3nX)$ and $(\gamma,\,4nX)$ reactions on $^{197}$Au in comparison with the only existing data set, which is the $(\gamma,\,3nX)$ one of Veyssiere~\emph{et al.}~\cite{veyssiere_1970}. The present $(\gamma,\,3nX)$ results are in good agreement with the Saclay ones on the increasing slope of the excitation function, which is the limited energy interval on which existing data are available. 

\subsection{Average photoneutron energies}

\begin{figure}[t]
\centering
\includegraphics[width=0.94\columnwidth, angle=0]{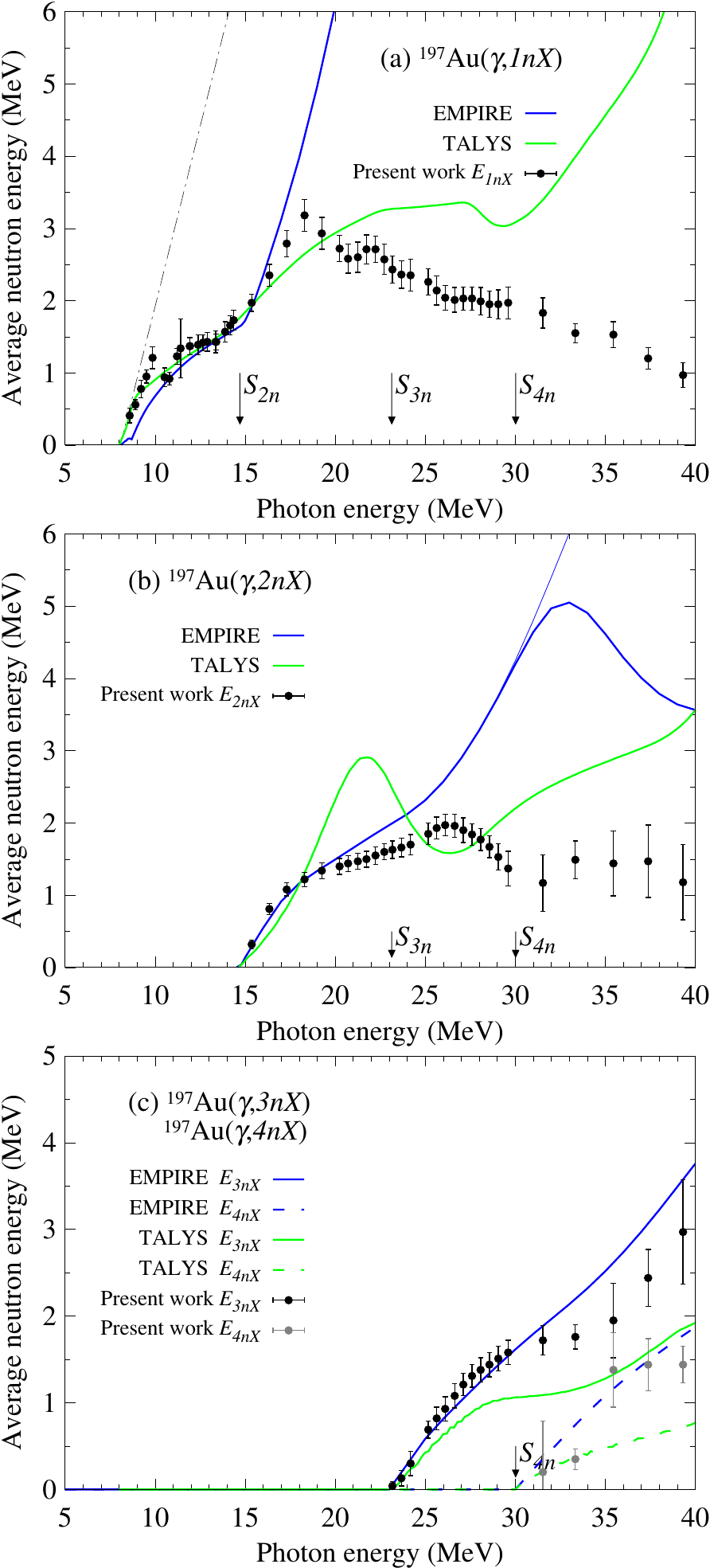} 
\caption{Present photoneutron average energies for the (a)~$(\gamma,\,1nX)$, (b)~$(\gamma,\,2nX)$, (c)~$(\gamma,\,3nX)$ and $(\gamma,\,4nX)$ reactions in $^{197}$Au compared with EMPIRE and TALYS calculations. The thin blue line in (b) shows EMPIRE $(\gamma,\,2n)$ average energies considering no charged particle emission.}\label{fig08_nen_gxn}     
\end{figure}

Figure~\ref{fig08_nen_gxn} shows the mean energies of the photoneutron spectra emitted in (a) $(\gamma,\,1nX)$, (b) $(\gamma,\,2nX)$ and (c) $(\gamma,\,3nX)$ and $(\gamma,\,4nX)$ on $^{197}$Au. 

The energy of the $(\gamma,\,1nX)$ photoneutrons shows a steep increase in the close vicinity of $S_n$, following closely the $(196/197)(E_\gamma-S_n)$ maximum kinetic energy given by the two body breakup kinematics. This suggests a non-statistic neutron emission in the $(\gamma,\,n)$ reaction in the low excitation energy region between $S_n$ and 10~MeV, a behavior which was also observed in the $(\gamma,\,n)$ reactions on $^{209}$Bi~\cite{gheorghe_2017} and $^{208}$Pb~\cite{gheorghe_2024}. The non-statistical energy increase is followed by a sharp $\sim$0.5~MeV drop at 11~MeV. Above it, the $E_{1nX}$ shows an increasing trend characteristic for statistical emission~\cite{filipescu_2024} which continues up to 18~MeV excitation energy, where the average energy of the $(\gamma,\,1nX)$ photoneutrons starts a slow and continuous decrease for entire remaining investigated energy range up to 39~MeV. 

The $(\gamma,\,2nX)$ reaction is characterized by a long rise in the average photoneutron energy $E_{2nX}$, which increases quickly above $S_{2n}$. The growth slows down at $\sim$20~MeV excitation energy and continues up to 27~MeV, above which the experimental $E_{2nX}$ shows a 0.5~MeV decrease followed by a stabilization on a 1~--~1.5~MeV plateau at excitation energies above 30~MeV. The $E_{3nX}$ and $E_{4nX}$ average neutron energies show increasing trends on the entire investigated energy range, from the $S_{3n}$ and $S_{4n}$ thresholds, respectively, and up to 39~MeV. 

\section{Statistical model calculations} \label{sec_STAT_calc}

The present photonuclear experimental data for $^{197}$Au are now compared with statistical model calculations obtained by the EMPIRE \cite{herman_2007_empire} and TALYS \cite{koning_2023_talys} codes. The first step in the theoretical modeling was the reproduction of the entrance channel described in both reaction codes by the cumulative contribution of the GDR excitation dominant at 10 to 30 MeV energies, and, at higher energies, of the photoabsorption on a neutron-proton pair (a quasideuteron, QD). The fit to experimental photoabsorption cross sections was tested with several Lorentzian-type closed-forms (SLO, MLO1, SMLO) of the GDR contribution for the E1 $\gamma$-ray strength functions~\cite{capote2009_ripl,plujko_2018}. 

\begin{figure}[t]
\centering
\includegraphics[width=0.98\columnwidth, angle=0]{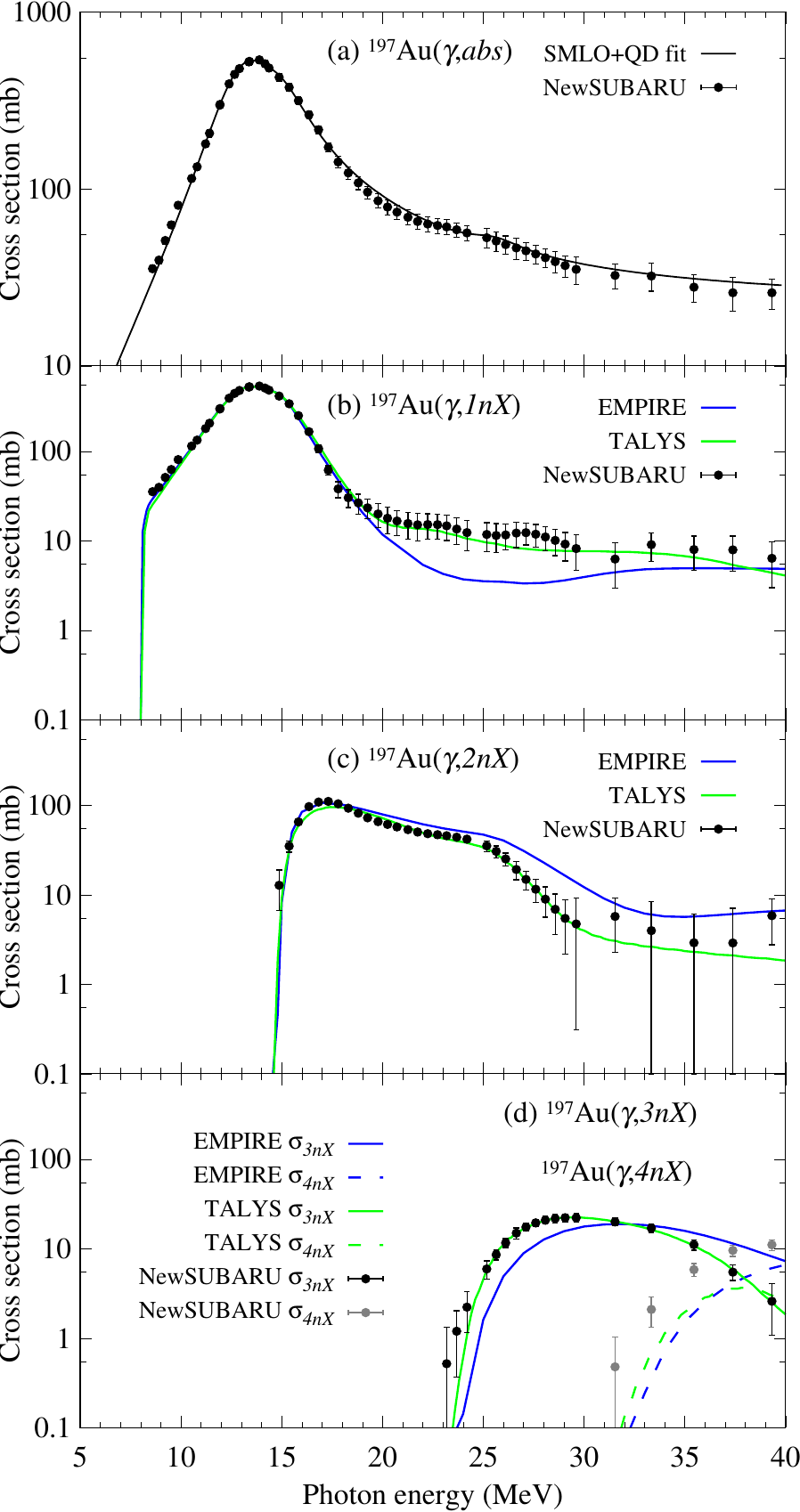} 
\caption{Present cross sections (filled circles) for the (a)~$(\gamma,\,abs)$, (b)~$(\gamma,\,1nX)$, (c)~$(\gamma,\,2nX)$, (d)~$(\gamma,\,3nX)$ and $(\gamma,\,4nX)$ reactions in $^{197}$Au compared with EMPIRE and TALYS statistical model calculations. The SMLO+QD curve in (a) corresponds to a three-Lorentzian fit to the present photoabsorption data using the Simple Modified Lorentzian function described in Ref.~\cite{plujko_2018,goriely_2019} and a quasideuteron (QD) component.}\label{fig_cs_calc}     
\end{figure}

We found that the experimental photoabsorption data were best reproduced by describing the $\gamma$SF by a sum of three SMLO-type Lorentzians plus a standard QD contribution. Table~\ref{tab_gdr} gives the values for the peak cross section, centroid energy, and width of the three SMLO-type Lorentzians. The resulting SMLO and QD component fit of the photoabsorption cross section is shown in Fig.~\ref{fig_cs_calc}(a).

The structure at 25 MeV can be identified with the isovector giant quadrupole resonance (IVGQR), as it has been previously done in Ref.~\cite{gordon_1977}. Using the approximate expression for the energy of $\Delta$T=1 quadrupole vibrations~\cite{bohr_mottelson} one obtains the location of the IVGQR at 23.2 MeV. Using a more detailed calculation~\cite{ishkhanov_2013} one can estimate the integrated IVGQR cross section at 267 mb$\cdot$MeV and the width $\Gamma$ = 3.6 MeV, which is consistent with the third resonance given in Table~\ref{tab_gdr}, assumed here to be of a dipole nature.

\begin{table}[b]
\caption{\label{tab_gdr} $^{197}$Au dipole resonance parameters adopted within the SMLO model for the three Lorentzians. $\sigma_0$ corresponds to the peak cross section, $E_0$ to the centroid energy and $\Gamma$ to the full width at half maximum. $i$ is the Lorentzian index.}
\begin{ruledtabular}
\begin{tabular}{lccr}
$i$ & $\sigma_{0}^{(i)}$ (mb) & $E_{0}^{(i)}$ [MeV] & $\Gamma^{(i)}$ [MeV] \\ \colrule
1   & 483.14                        & 13.50                     & 3.94                          \\
2   & 116.22                        & 15.08                     & 2.82                          \\
3   & 6.36                          & 25.67                     & 3.04                          \\ 
\end{tabular}
\end{ruledtabular}
\end{table}

Since the EMPIRE code can only consider a single OMP common to all nuclei in the reaction chain, the Koning-Delaroche general spherical OMP (RIPL ID - 2405) is used in the EMPIRE calculations. The level densities were described with the enhanced generalized superfluid model \cite{herman_2007_empire} in the EMPIRE code. In contrast, the TALYS calculations make use of microscopic mean-field-based models for the nuclear level density based on the temperature-dependent Hartree-Fock-Bogolyubov (HFB) plus combinatorial model \cite{hilaire_2012} and the JLMB optical mode potential  \cite{bauge_2001} adopted for the target and residual nuclei. 

In EMPIRE, the level densities have been adjusted to reproduce the low-lying discrete level scheme of all nuclei involved in the neutron chain (note that no $s$-wave resonance spacing exist for the Au isotopes considered here). To reproduce the present experimental results, 
we found it necessary in the EMPIRE calculations to reduce the so determined $^{196}$Au level density by lowering the $\tilde{a}$ parameter by 10\% and also to increase the first neutron preequilibrium contribution computed within the PCROSS model~\cite{capote_1991_pcross}. This increased the $(\gamma,\,1nX)$ cross section tail and lowered the $E_{2nX}$ average neutron energies at excitation energies above 30~MeV. 
With TALYS, both the total and partial particle-hole level densities have been adjusted to optimize the description of the three cross sections  $\sigma_{\gamma,inX}$ ($i=1, 2, 3$).

Figures~\ref{fig_cs_calc}(b)--(d) show the present photoneutron cross sections compared with EMPIRE and TALYS calculations. The $^{197}$Au$(\gamma,1nX)$ cross section is overall well described by the EMPIRE calculations except for the 20 to 30~MeV excitation energy region, where the calculations underestimate the experimental values. The EMPIRE calculations tend to overestimate the $^{197}$Au$(\gamma,2nX)$ cross section for excitation energies above 20~MeV. Both the $(\gamma,3nX)$ and the $(\gamma,4nX)$ are underestimated by the EMPIRE calculations. TALYS description of the $^{197}$Au$(\gamma,1nX)$ cross section is compatible with the new data but does not show the increased values around 27~MeV. The $(\gamma,2nX)$ and $(\gamma,3nX)$ cross sections are also in excellent agreement with data, while the $(\gamma,4nX)$ cross section is underestimated.

The present experimental average photoneutron energies are compared in Fig.~\ref{fig08_nen_gxn} with EMPIRE and TALYS calculations. Here, we plot the average energies of the exclusive neutron spectra obtained in the EMPIRE and TALYS calculations, which contain the summed contributions of reactions with emission of a given $i$ neutron multiplicity. The EMPIRE and TALYS model calculations describe qualitatively well the $E_{inX}$ behavior at the low incident photon energies in the vicinity of the corresponding reaction thresholds. However, the EMPIRE calculations couldn't reproduce the high energy $(\gamma,\,n)$ neutron emission just above $S_n$ and neither the structure observed at 10.5~MeV excitation energy. Also, the long decrease in the $E_{1nX}$ energy starting at 18~MeV excitation energy could not be reproduced by the EMPIRE nor TALYS calculations. Although the EMPIRE calculations follow closely the increasing $E_{2nX}$ experimental values at low incident energies, the calculations fail to reproduce the average energy decrease and its stabilization on a 1~--~1.5~MeV plateau at excitation energies above 25~MeV. A decrease in the $E_{2nX}$ is predicted in the EMPIRE calculations at incident energies of 33~MeV, but this follows a continuous increase in average neutron energies up to 5 MeV, well above experimental values. The EMPIRE predicted $E_{2nX}$ drop is due to the contribution of charged particle emission channels, which can be seen from the continuous increasing trend of the average energies of neutrons emitted in the $(\gamma,\,2n)$ reaction, represented by thin solid line in Fig.~\ref{fig08_nen_gxn}(b). TALYS description also fails for this channel, the energy dependence being in disagreement with data for almost the entire range. 
While EMPIRE's description of the $(\gamma,\,3n)$ and $(\gamma,\,4n)$ average energies are rather satisfactorily described, TALYS tends to underestimate them.

\section{Summary and conclusions} \label{sec_summary} 

Using quasi-monochromatic LCS $\gamma$-ray beams and a high-and-flat efficiency neutron detection system, we performed photoneutron cross section measurements on the $^{197}$Au nucleus in and above the GDR region. We provide a complete set of $(\gamma,\,1-4nX)$ photoneutron cross sections and average photoneutron energies for the excitation energy range between 8 and 39~MeV. The total photoabsorption cross section was extracted as the sum of the photoneutron ones. The GDR parameter values were obtained by fitting of theoretical photoabsorption cross sections to the present experimental data.  

There is good agreement in the GDR region between the present photoabsorption cross sections and the Saclay results of Veyssiere~\emph{et al.}~\cite{veyssiere_1970}. Thus, the results of the present study do not support the recommendation of Berman~\emph{et al.}~\cite{berman_1987} of lowering the Saclay photoabsorption cross sections by 8$\%$.

A non-statistical high-energy neutron emission was observed for the $(\gamma,\,n)$ reaction in the low-energy region between $S_n$ and 10~MeV, a behavior also displayed in our previous measurements on $^{209}$Bi~\cite{gheorghe_2017} and $^{208}$Pb~\cite{gheorghe_2024}. 

All the experimental results obtained in the present paper are available in numerical format in the Supplemental Material~\cite{supplemental_material}.

\section{Acknowledgments}
The authors are grateful to H. Ohgaki of the Institute of Advanced Energy, Kyoto University for making a large volume LaBr$_3$:Ce detector available for the experiment.   
I.G., D.F. and M.K. acknowledge the support from the Extreme Light Infrastructure Nuclear Physics (ELI-NP) Phase II, a project cofinanced by the Romanian Government and the European Union through the European Regional Development Fund - the Competitiveness Operational Programme (1/07.07.2016, COP, ID 1334).
I.G. and D.F. acknowledge the support from the Romanian Project PN-23-21-01-02.
S.G. acknowledges financial support from F.R.S.-FNRS (Belgium). This work was supported by the F.R.S.- FNRS and FWO under the EOS Project Nr O022818F.

\bibliographystyle{apsrev4-1}
\bibliography{newsubarubibfile}

\end{document}